\documentclass{article}
\usepackage{spconf,amsmath,color,url}
\usepackage{hyperref}
\usepackage{graphicx} 

\let\OLDthebibliography\thebibliography
\renewcommand\thebibliography[1]{
  \OLDthebibliography{#1}
  \setlength{\parskip}{0pt}
  \setlength{\itemsep}{0pt plus 0.3ex}
}

\ninept

\begin{document}\sloppy

\def\x{{\mathbf x}}
\def\L{{\cal L}}

\newcommand{\comment}[1]{\textcolor{blue}{#1}} 

\title{The 2020 ESPnet update: new features, broadened applications, performance improvements, and future plans}
%

\name{
\begin{tabular}{c}
\it Shinji Watanabe$^{1}$, Florian Boyer$^{2,3}$, Xuankai Chang$^1$, Pengcheng Guo$^{4,1}$, Tomoki Hayashi${^{5,6}}$\\
\it Yosuke Higuchi$^7$, Takaaki Hori$^8$, Wen-Chin Huang$^6$, Hirofumi Inaguma${^9}$, Naoyuki Kamo$^{10}$, \\
\it Shigeki Karita$^{11}$, Chenda Li$^{12}$, Jing Shi$^{13}$, Aswin Shanmugam Subramanian$^{1}$, Wangyou Zhang$^{12}$
\end{tabular}
}

\address{
    $^1$Johns Hopkins University, 
    $^2$Airudit, Speech Lab., $^3$LaBRI, Bordeaux INP, CNRS, UMR 5800, 
    \\
    $^4$Northwestern Polytechnical University, $^5$Human Dataware Lab. Co., Ltd.,
    $^6$Nagoya University 
    \\
    $^7$Waseda University, 
    $^8$MERL,
    $^9$Kyoto University, 
    $^{10}$NTT Corporation,
    $^{11}$Google\\
    $^{12}$Shanghai Jiao Tong University,
    $^{13}$Institute of Automation, Chinese Academy of Sciences \\
}

\maketitle

\begin{abstract}
This paper describes the recent development of ESPnet (\url{https://github.com/espnet/espnet}), an end-to-end speech processing toolkit.
This project was initiated in December 2017 to mainly deal with end-to-end speech recognition experiments based on sequence-to-sequence modeling.
The project has grown rapidly and now covers a wide range of speech processing applications.
Now ESPnet also includes text to speech (TTS), voice conversation (VC), speech translation (ST), and speech enhancement (SE) with support for beamforming, speech separation, denoising, and dereverberation.
All applications are trained in an end-to-end manner, thanks to the generic sequence to sequence modeling properties, and they can be further integrated and jointly optimized.
Also, ESPnet provides reproducible all-in-one recipes for these applications with state-of-the-art performance in various benchmarks by incorporating transformer, advanced data augmentation, and conformer.
This project aims to provide up-to-date speech processing experience to the community so that researchers in academia and various industry scales can develop their technologies collaboratively.
\end{abstract}
\section{Introduction}
\label{sec:intro}
The rapid growth of deep learning techniques has made significant changes and improvements in various speech processing algorithms.
Automatic speech recognition (ASR) is one of the successful examples in this trend, which achieved significant performance gains with a hybrid model based on hidden Markov model (HMM) and deep neural network (DNN)~\cite{hinton2012deep}.
The emergent sequence to sequence techniques further accelerate this trend ~\cite{cho2014learning,sutskever2014sequence}.
Thus, we realized end-to-end neural ASR modeling based on these sequence to sequence techniques~\cite{graves2013speech,chorowski2015attention,chan2016listen}.

Due to the significant demand to establish end-to-end ASR and other speech processing applications, we started developing ESPnet, an end-to-end speech processing toolkit, in December 2017.
Our original implementation followed the success of Kaldi speech recognition toolkit~\cite{povey2011kaldi} and leveraged deep learning frameworks based on Chainer~\cite{tokui2015chainer} and PyTorch~\cite{NEURIPS2019_9015}.
This paper introduces the recent advances of the ESPnet project since our previous official publication of the whole ESPnet project in 2018 \cite{watanabe2018espnet}.

\subsection{Related framework}
There are a number of excellent deep learning frameworks that realize similar functions to what ESPnet covers, e.g., Fairseq~\cite{ott2019fairseq}, RETURNN~\cite{zeyer2018returnn}, Lingvo~\cite{shen2019lingvo}, and NeMo~\cite{kuchaiev2019nemo}.
These frameworks provide many AI applications, including various natural language processing (NLP) and speech processing methods, based on sequence-to-sequence modeling.
Most frameworks include ASR, Text-to-Speech (TTS), and neural machine translation or speech translation (ST).
Compared with them, ESPnet focuses more on a wide range of speech applications, and in addition to the above applications, ESPnet also supports various speech enhancement functions now.
Also, there are numbers of specialized toolkits for each speech application including Wav2Letter~\cite{collobert2016wav2letter} and Espresso~\cite{wang2019espresso} for ASR, Fairseq-S2T~\cite{wang2020fairseq} for ST, and Asteroid~\cite{pariente2020asteroid} for speech separation.
The above activities are complementary, and we are closely collaborating/interacting with them, especially with Facebook torchaudio, Nivida Nemo, and Asteroid teams.

\subsection{Development history since 2018}

\paragraph*{Broadened applications}
The most remarkable update in the recent advances of ESPnet is to broaden the applications from ASR tasks.
TTS and Voice conversion (VC) are the most notable extensions, which will be discussed in Section \ref{sec:espnet-tts} as ESPnet-TTS.
Note that ESPnet-TTS is also designed to integrate with ASR/TTS joint training \cite{hori2019cycle} as one of the core research topics in the Fifth Frederick Jelinek Memorial Summer Workshop.
Speech translation was also developed to support our research activities to simplify complex speech translation studies with reproducible all-in-one recipes, which will be discussed in Section \ref{sec:espnet-st}.
Section \ref{sec:espnet-se} describes one of the most recent significant updates in terms of the applications by involving speech enhancement, which includes beamforming, speech separation, denoising, and dereverberation.
In addition to these major speech applications, ESPnet also include CTC-based voice activity detection \cite{yoshimura2020end} and CTC-based force alignment \cite{kurzinger2020ctc}.
\vspace*{-3mm}
\paragraph*{Notable neural architectures and learning methods}
ESPnet has promptly followed various techniques to provide state-of-the-art results for the community.
For example, we have been developing the several  ASR decoding algorithms, as described in Sections \ref{sec:advanced-decoding}, \ref{sec:generalized-decoding}, and \ref{ssec:nar_asr}, which aim to realize fast, unified, and parallelizable inference algorithms.
Transformer, conformer, and data augmentation techniques, as described in Sections \ref{ssec:transformer} and \ref{ssec:conformer} significantly improve the ASR performance, and they are also used in other ESPnet applications.
Section \ref{ssec:rnnt} also describes RNN-Transducer, which is another crucial example aimed for on-line/streaming ASR.







\vspace*{-3mm}
\paragraph*{New DNN training system based on ESPnet2}
In the release of ESPnet v.0.7.0, we created a new system for DNN training to extend our system for future developments. We referred this project as ESPnet2. Although the main purpose of ESPnet2 is refactoring, it is not limited to refining the source code only. We added several new/enhanced features, including distributed training, on-the-fly feature extraction from the raw waveform, and solving a memory allocation issue of original ESPnet when training on a large scale corpus. The training system of ESPnet2 is shared with all DNN tasks, ASR, TTS, SE, etc., and we can easily integrate a new task with this system. We have already migrated ASR and TTS parts to ESPnet2 and also planning to migrate all existing features in the next steps.


%



In addition to the above applications, techniques, and ESPnet2-based new training system, we have also improved software workflow by enhancing the continuous integration, enriching documentation, supporting the docker, pip install, and model zoo functions.

\subsection{Activity statistics}
Table \ref{table:activity_statistics} discusses the activity statistics about ESPnet between August 2018 (when~\cite{watanabe2018espnet} was presented at Interspeech 2018), December 2019, and the current one (December 2020).
We obtained the previous GitHub statistics from \url{web.archive.org}\footnote{We obtained the GitHub statistics from \url{http://web.archive.org/web/20180828121301/https://github.com/espnet/espnet} and \url{http://web.archive.org/web/20191213095813/https://github.com/espnet/espnet}}.
The citation counts were obtained from Google Scholar, accessed on December 10, 2020.
\begin{table*}[tbh]
\centering
\scalebox{0.90}{
\begin{tabular}{r||c|cccccc|c}
                    & Citations & Contributors & Watches & Stars & Forks & Commits & Issues/PRs & Recipes \\
                    \hline
Aug. 2018 (v.0.2.0) & 19        & 17           & 57      & 610   & 141   & 1,216    & 380         & 19      \\
Dec. 2019 (v.0.6.0) & 95        & 54          & 114     & 1.7K  & 506   & 5,994    & 1,484       & 47   \\
Dec. 2020 (v.0.9.6) & 321       & 94           & 160     & 3.1K  & 976   & 10,080   & 2,768       & 62 (27)  
\end{tabular}
}
\caption{Activity statistics, including GitHub development statistics, citation counts, and the numbers of supported recipes.
The numbers of recipes in December 2020 include 62 recipes in ESPnet1 and 27 recipes in ESPnet2.
}
\label{table:activity_statistics}
\vspace*{-3mm}
\end{table*}
Table \ref{table:activity_statistics} shows that all GitHub development statistics and citation counts have significantly increased in these two years.
In terms of the GitHub development statistics, the large increase in the numbers of contributors, forks, and commits shows that many active developers have increasingly supported the development of ESPnet.
The number of citations has also been growing from 19 in 2018 to 95 in 2019, and 205 in 2020.
This citation count shows that ESPnet has been used in various research groups and contributed a lot to speech research activities.
Finally, we also list the numbers of supported recipes in these periods.
We also started to release the ESPnet2 recipes on May 24, 2020, and it includes 27 recipes already.


\section{ESPnet-ASR}\label{ssec:espnet-asr}


\subsection{Advanced decoding}\label{sec:advanced-decoding}
The basic decoding algorithm in ESPnet-ASR follows an output-label synchronous beam search, which efficiently finds the most probable label sequence for a given input utterance using an encoder-decoder (RNN, Transformer, etc.), CTC and a language model (LM)~\cite{hori2017advances}.
CTC guides the beam search process to keep valid hypotheses with monotonic alignments, excluding those with irrelevant alignments. The LM provides a significant improvement of accuracy when it is trained with a large amount of in-domain text.

The decoding module also includes the following features:
\vspace*{-4.5mm}
\paragraph*{Accelerated decoding}
ESPnet accelerates the decoding process by vectorizing multiple hypotheses during the beam search~\cite{seki2019vectorized}, where the score accumulation steps for each hypothesis are implemented as vector-matrix operations for the vectorized hypotheses. This strategy allows us to take advantage of the parallel computing capabilities of multi-core CPUs and GPUs, resulting in significant speedups. This manner is also applied to the scoring steps by the CTC and LM to reduce the overhead.
\vspace*{-3mm}
\paragraph*{Use of word-based LM}
ESPnet typically employs an encoder-decoder and an LM relying on a common label set for the sake of simplicity in decoding~\cite{hori2017advances}, but it also allows us to use such models while relying on different label sets. For example, we can combine a word-based LM on top of the encoder-decoder that emits letters for English ASR. This kind of configuration is useful for the case that a sufficient amount of paired data is not available to train an encoder-decoder that emits labels longer than letters, and a large amount of external text is available to reliably train the word-based LM. Currently, ESPnet supports multi-level LM~\cite{hori2017multi} and look-ahead LM~\cite{hori2018end} as a framework to incorporate a word-based LM into the decoding process.

\subsection{Generalized decoding}
\label{sec:generalized-decoding}
Recent ESPnet models inherit an abstract class named \texttt{ScorerInterface} for model-agnostic beam search decoding.
It receives encoder output sequences, hypotheses, and internal states (i.e., cache) as inputs, and returns scores of decoder outputs to append to the hypotheses.
Most of the models, for example, RNN ASR models, and LM implement this unified interface. In decoding, we combine them as one PyTorch module that can easily run in CPUs and GPUs as well as fairseq~\cite{ott2019fairseq}'s search module\footnote{\url{https://github.com/pytorch/fairseq/blob/master/fairseq/search.py}}.
Similar to Lingvo~\cite{shen2019lingvo}'s beam search\footnote{\url{https://github.com/tensorflow/lingvo/blob/master/lingvo/core/beam_search_helper.py}}, we support efficient joint scoring~\cite{ESPnetIEEESP2017} with CTC and n-gram LM (e.g., KenLM~\cite{kenlm-heafield-etal-2013-scalable}) by a specialized class named \texttt{PartialScorerInterface}. It scores a subset of hypotheses with higher scores in the preceded faster models.

\subsection{Transformer}\label{ssec:transformer}

\begin{figure}
	\centering
    \includegraphics[width=\columnwidth]{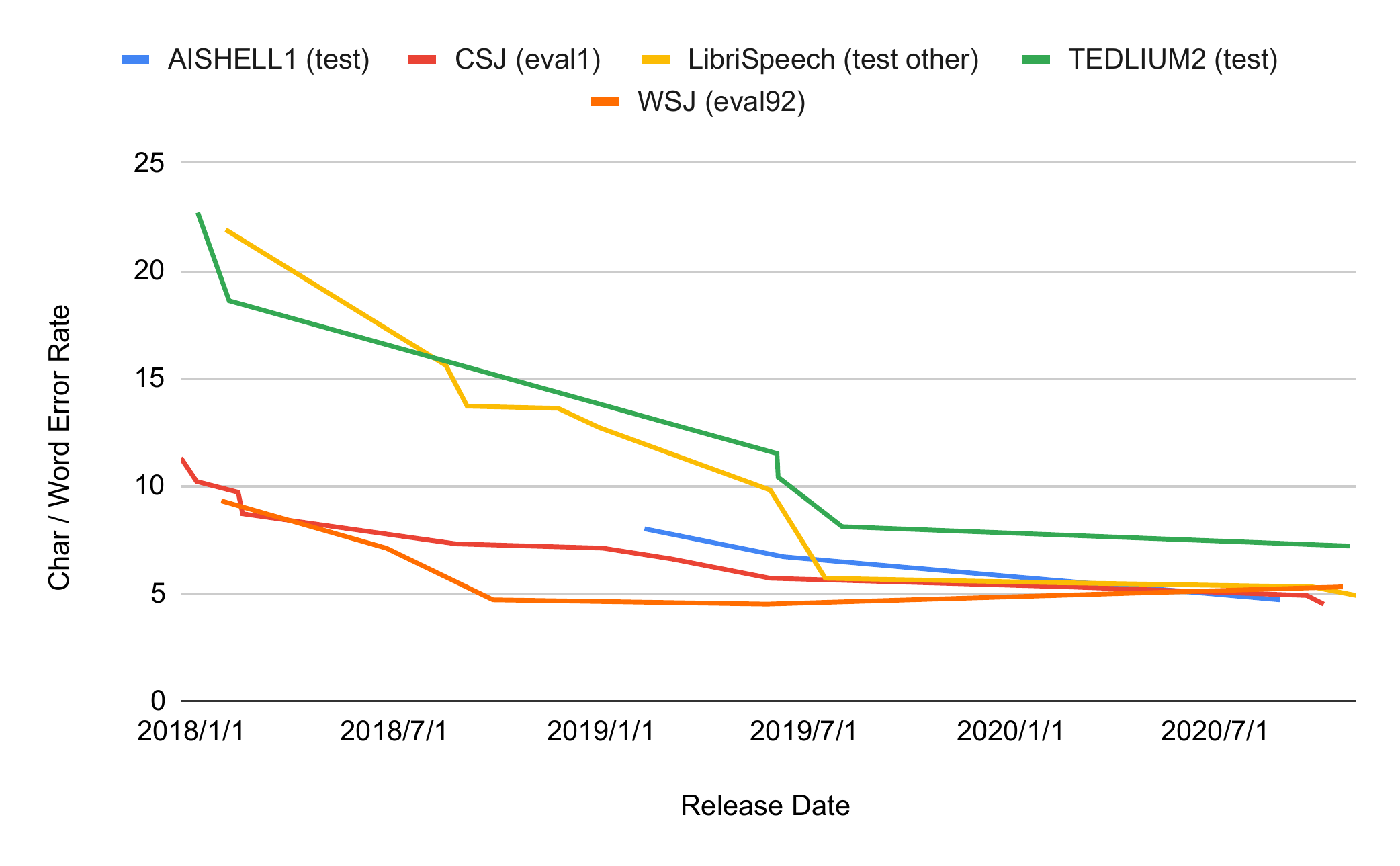}
    \vspace*{-9mm}
    \caption{A history of char/word error rates (CER/WER) on various ASR tasks in ESPnet. 
    }
    \label{fig:asr}
    \vspace*{-2mm}
\end{figure}


Figure~\ref{fig:asr} illustrates a history of reported char/word error rates (CER/WER) on major ASR corpora in ESPnet.
In Figure~\ref{fig:asr}, we can observe some large reductions in 2019 by Transformer-based ASR and data augmentation~\cite{ESPnetASRU2019}.
The Transformer~\cite{transformer} is a feed-forward architecture using self-attention mechanism for sequence modeling.
One drawback in the Transformer is slow decoding speed because it transforms an entire sequence to decode every single token, while the RNN requires just one recurrent frame.
To address this problem, the \texttt{ScorerInterface} caches its previous attention matrix as a state, and computes only new outer vectors to append to the matrix for each step. 
This method accelerates not only our Transformer-based ASR but also TTS inference as fast as RNN.


\subsection{Conformer}\label{ssec:conformer}
In Figure~\ref{fig:asr}, we can observe further improvement in 2020 by Conformer-based ASR~\cite{guo2020recent}. While Transformer models can learn long-range global context better than RNN models, they are less capable to exploit the local information. To address this drawback, Gulati et al.~\cite{gulati2020conformer} proposed to combine both self-attention and convolution in the encoder, which is named Conformer encoder. With this design, the self-attention module captures the global context while the convolution module exploits the local correlations synchronously. By integrating the Conformer encoder with a Transformer decoder, our models achieved WER of 4.9\% on the LibriSpeech test\_other task, CER of 4.7\% on the AISHELL1 test task. Besides, we also obtained a 7\% relative improvement on the multi-speaker WSJ-2mix data and a more than 15\% relative improvement on 8 low-resource language corpora compared with Transformer models~\cite{guo2020recent}.

\subsection{RNN-Transducer}\label{ssec:rnnt}

Alongside CTC, attention and hybrid models, ESPnet also supports models based on the RNN-T loss proposed by A. Graves \cite{Graves12-STW}. Following recent advances, different encoder-decoder architecture are also available for these models such as RNN, Transformer (\ref{ssec:transformer}) and Conformer (\ref{ssec:conformer}), but also \textit{free-form} architecture. The latter, exclusive to transducer-based models in ESPnet, allows previously described neural networks and additional ones (TDNN and Causal-Conv1d \cite{Weng19-MBR}) to be combined freely to form a \textit{customized} encoder and decoder architecture definition\footnote{\url{https://espnet.github.io/espnet/tutorial.html\#transducer}}.
For inference with transducer-based models, we follow a different decoding procedure than \ref{sec:advanced-decoding} but share the interface described in \ref{sec:generalized-decoding}. Here, we propose four decoding algorithms allowing more flexibility in regards to the performance-speed trade-off: greedy decoding constrained to one expansion step, beam search \cite{Graves12-STW} (without prefix search), time-synchronous decoding \cite{Saon20-ALS}, alignment-length decoding \cite{Saon20-ALS} and n-step constrained beam search with our modified version of one-step constrained beam search \cite{Juantae20-ART}. Shallow fusions with RNN-based and Transformer-based LM and multi-level LM decoding are also supported (\ref{sec:advanced-decoding}).

At the time of writing, we made available in ESPnet various training and decoding configurations and report results for a few corpora: VIVOS (Vietnamese), Voxforge (Italian), Commonvoice (Czech and Welsh) and plan to expand the coverage to other corpora in the following months. Additionally, transducer-based models from ESPnet were reported successful by various researchers with other corpora and languages: OpenSTT (Russian) \cite{Andrusenko20-EEE}, ESTER (French) \cite{Boyer19-ETE} and CHiME-6 Challenge \cite{Andrusenko20-TCE}.

\subsection{Non-autoregressive modeling}\label{ssec:nar_asr}
In addition to the autoregressive (AR) modeling based on an encoder-decoder architecture~\cite{chorowski2015attention, chan2016listen}, 
ESPnet supports non-autoregressive (NAR) modeling of end-to-end ASR models.
AR models suffer from slow inference speed, which requires as many forward calculations of the decoder as the length of an output sequence.
On the other hand, NAR models permit for fast sequence generation with a constant number of the decoder calculations.
Recently proposed Mask-CTC~\cite{higuchi2020mask} realizes the NAR modeling based on 
mask-predict~\cite{ghazvininejad2019mask} and CTC. 
During inference, a target sequence is generated by iteratively refining an output of CTC with mask-predict.
Mask-CTC achieves fast inference time ($<$ 0.1 RTF using a single CPU) and 
competitive performance to that of the AR model using Conformer (Section \ref{ssec:conformer})~\cite{higuchi2020improved}.

\section{ESPnet-TTS}\label{sec:espnet-tts}

\subsection{TTS}\label{ssec:tts}

As an extension of ESPnet-ASR, we released ESPnet-TTS that supports text-to-speech (TTS) task~\cite{hayashi2020espnet}, which mainly focuses on the development of text to mel-spectrogram (text2mel) models.
Initially, ESPnet-TTS has been developed with ESPnet1, but recently it is also available in ESPnet2. 
It supports AR text2mel models such as Tacotron~2~\cite{shen2017tacotron2} and Transformer-TTS~\cite{li2018transformer}, NAR models such as FastSpeech~\cite{ren2019fastspeech} and FastSpeech~2~\cite{ren2020fastspeech}, and their multi-speaker extensions with X-vector~\cite{snyder2018x} and global style token~\cite{wang2018style}.
Thanks to the unified libraries with various tasks, it can quickly introduce new architectures from the other tasks into TTS  (e.g., Conformer-based FastSpeech).
As mel-spectrogram to waveform (mel2wav, i.e., vocoder) models, we support WaveNet vocoder~\cite{oord2016wavenet}, Parallel WaveGAN~\cite{yamamoto2020parallel}, MelGAN~\cite{kumar2019melgan}, and Multi-band MelGAN~\cite{yang2020multi} with external libraries\footnote{\url{https://github.com/r9y9/wavenet_vocoder}}\footnote{\url{https://github.com/kan-bayashi/ParallelWaveGAN}}. 
In addition that users can quickly develop the state-of-the-art baseline systems for the research purpose,  they can easily make a demonstration system, which works in real-time for various languages, including English, Mandarin, and Japanese\footnote{The realtime demo is available at \url{https://bit.ly/3a54G3s}.}.

\subsection{VC}\label{ssec:voice_conversion}

Considering that TTS and VC share the same goal of performing speech synthesis, it is easy to extend the TTS models into VC models. We recently released VC recipes that can be trained with a parallel corpus (i.e. pair of sentences with the same content uttered by the source and target speaker) for mel-spectrogram-to-mel-spectrogram conversion. We support Tacotron2-based and Transformer-based models, and we provided a pre-trained model that enables sample efficient training such that only approximately 5 minutes of data is needed \cite{Huang2020vtn}. In addition, an any-to-one system that chains an ASR and a TTS model was developed as the baseline system \cite{Huang2020asrtts} of the voice conversion challenge 2020 \cite{vcc2020}. Taking advantage of the well-tuned pre-trained ASR and TTS models provided in ESPnet, this simple yet effective system was placed 2nd in terms of conversion similarity in the official listening test.

\section{ESPnet-ST}\label{sec:espnet-st}
Although ESPnet has supported ASR and TTS tasks as initial speech applications, we recently started to support the speech translation (ST) task~\cite{inaguma-etal-2020-espnet}.
Both the traditional pipeline approach, where ASR and text-based machine translation (MT) modules are cascaded, and end-to-end (E2E) approach, where source speech is directly translated to text in another language, are readily available in ESPnet-ST.
Our goal is to build reliable baselines quickly with minimal effort, and we demonstrated the state-of-the-art translation performance~\cite{inaguma-etal-2020-espnet}.
Moreover, new progresses in ASR can be easily transferred to ST performance improvement since both tasks are seamlessly integrated into a unified ESPnet framework.
For instance, we presented that a better encoder architecture which was originally proposed in the ASR task, Conformer (Section \ref{ssec:conformer}), is effective for the E2E-ST task as well in~\cite{guo2020recent}.
Multilinguality is also a good example in the ST task~\cite{inaguma2019multilingual}.
To speed up inference, non-autoregressive (NAR) E2E-ST models have been studied on ESPnet-ST recently~\cite{inaguma2020orthros}, similar to NAR ASR in Section \ref{ssec:nar_asr}.
We will officially support various NAR methods in the future.

\section{ESPnet-SE}\label{sec:espnet-se}
ESPnet-SE \cite{li2020espnet} is a recently released component of ESPnet2. 
It has the capability to process various speech enhancement tasks, including speech dereverberation, denoising and speech separation for both single-channel and multi-channel data.
ESPnet-SE is designed to be easily integrated with the downstream ASR tasks, including the robust ASR and multi-talker ASR.
The enhancement front-end can be also jointly trained with ASR tasks.

\subsection{Single-channel enhancement/separation}
\label{ssec:sin-chan-sep}
In the single-channel condition, we have implemented various popular models for speech enhancement and separation.
Both frequency-domain \cite{yu2017permutation,kolbaek2017multitalker} models and time-domain \cite{luo2018tasnet,luo2019conv} models are supported.
The networks and loss functions used in the speech enhancement/separation are flexible with configuration files, thus the newest models can be quickly reproduced and included in the toolkit.
The input and output of the SE model interface are raw waveforms, which makes it easier for the integration with other downstream tasks.
Depending on the number of speech sources, single-talker speech enhancement and multi-talker speech separation are designed into a unified framework in ESPnet-SE.

\subsection{Multi-channel enhancement/separation}
\label{ssec:multi-chan-sep}
When processing multi-channel speech data collected by multiple microphones or a microphone array, the additional spatial information can be exploited to achieve better performance with ESPnet-SE.
The current model \texttt{BeamformerNet} is mainly based on neural beamformers~\cite{Neural-Heymann2016, Improved-Erdogan2016}, with support for various beamformer types including minimum variance distortionless response (MVDR)~\cite{Beamforming-Van1988}, weighted power minimization distortionless response (WPD)~\cite{Unified-Nakatani2019}, and weighted minimum power distortionless response (wMPDR)~\cite{Jointly-Boeddeker2020}.
It is worth noting that \texttt{BeamformerNet} is implemented in a fully differentiable way, thus allowing end-to-end optimization with either signal-level objectives or the ASR criterion.

\subsection{Dereverberation}
\label{ssec:dereverb}
Support for DNN based weighted prediction error (WPE) dereverberation \cite{Neural-Kinoshita2017} is added \footnote{DNN-WPE module from \url{https://github.com/nttcslab-sp/dnn_wpe}}. The dereverberation network can be optimized using the target signal like \cite{Neural-Kinoshita2017}. We also provide an option to connect it as a dereverberation subnetwork before the differential beamformer described  in Sec.~\ref{ssec:multi-chan-sep} and train it end-to-end with just speech recognition objectives as used in \cite{aswin_waspaa2019}. Evaluation is performed with all the speech enhancement metrics prescribed in the REVERB challenge \cite{reverb_challenge} in addition to the ASR metrics.

\subsection{Multi-speaker ASR}
\label{ssec:multi-spkr-asr}
In addition to the multi-channel speech separation in Sec.~\ref{ssec:multi-chan-sep}, the end-to-end multi-channel multi-speaker speech recognition~\cite{chang2019mimo,chang2020end} is supported in ESPnet-SE. In line with the multi-channel speech separation, the input is the multi-channel speech signals containing overlapped speech from multiple speakers. The model can be divided into two modules, including speech separation and recognition. Firstly, the speech separation module is a masking network-based neural beamformer as introduced in Sec.~\ref{ssec:multi-chan-sep}. The masking network first separates the overlapped speech by predicting signal masks for each speaker and an additional noise mask over all channels. Then the beamformer enhances the speech of each speaker. Secondly, the speech recognition module, based on the joint CTC /attention-based E2E ASR framework~\cite{hori2017advances}, takes the separated speech of each speaker as input and outputs the corresponding transcriptions. All the parameters in the whole model are optimized simultaneously based on the final ASR loss only. To address the label permutation problem in computing the ASR loss, we compute the ASR loss with permutation invariant training (PIT), which is similar to the method used in previous end-to-end multi-speaker ASR systems~\cite{seki2018purely,chang2019end}.

\subsection{Joint training}
Thanks to the complete modules of both speech enhancement and recognition designed in an unified framework, the joint training of both the front-end model and the back-end model can be achieved without any hassle. Based on the foundation from single/multi-channel enhancement/separation in~Sec.~\ref{ssec:sin-chan-sep},~\ref{ssec:multi-chan-sep},~\ref{ssec:dereverb} and multi-speaker ASR in~\ref{ssec:multi-spkr-asr}, joint-training incorporates the models and functions from both, along with some specific features related to the joint-training, such as the composition of the total loss, the formation of the enhanced sources, the criterion to permute the sources and so on. In general, with a serial pipeline from enhancement to recognition, all the built-in models from Sec.~\ref{ssec:sin-chan-sep},~\ref{ssec:multi-chan-sep}, ~\ref{ssec:dereverb} and ~\ref{ssec:multi-spkr-asr} could be configurable, corresponding to the input data. The whole design of joint-training is end-to-end, and all the parameters in the whole model could be optimized with the combination of enhancement loss and ASR loss or just the final ASR loss. In terms of the application, jointly training models avoid the complexity from the individual training of the front-end and back-end, as well as the difficulties to combine these two parts in both training and inference phase. After the experiments on some benchmark datasets in multi-speaker scenes, e.g. WSJ0-2mix~\cite{2016Deepclustering}, the jointly trained models attain performance that is comparable to or even better than the baseline models.

\section{Summary and future plans}
This paper introduced the recent advances of the ESPnet end-to-end speech processing toolkit.
The ESPnet project has been rapidly growing by covering a wide range of speech applications and incorporating state-of-the-art techniques based on community-driven developments by many supporters.
Our future work will further accelerate the project by considering the demand of the community and research trends.
For example, we will focus on enhancing online/streaming ASR functions from our current RNN-T implementation and further broadening the speech applications by realizing end-to-end speech-to-speech translation and spoken dialogue systems within the framework. We will also focus on realizing complete speech conversation understanding systems that can recognize who spoke what, when, and where by incorporating models such as \cite{subramanian2020directional}.


\bibliographystyle{IEEEbib}
\bibliography{refs}

\end{document}